\documentclass[10pt,letterpaper,twocolumn]{article} 
\usepackage{ol2} 
\usepackage{hyperref,cite}

\usepackage{graphicx,ulem}
\usepackage{amsmath,amstext,amssymb}
\usepackage{bm}

\newcommand{\Aeff}[1]{A_{{\rm eff},#1}}%

\newcommand{\Pin}{P_{\rm in}}
\newcommand{\Aovl}{A_{\rm SHG}}

\newcommand{\nbar}{\bar n}%
\newcommand{\nshg}{N_{\rm SHG}}%
\newcommand{\nkerr}{N_{\rm Kerr}}%
\newcommand{\neff}[1]{n_{{\rm eff},#1}}%
\newcommand{\x}{\mathbf{x}}
\newcommand{\E}{\mathbf{E}}

\newcommand{\Es}{\mathcal{E}}
\newcommand{\chit}{\chi^{(2)}}
\newcommand{\kz}{\beta}

\newcommand{\bp}{\kz^{(1)}}

\newcommand{\bpp}{\kz^{(2)}}
\newcommand{\Dm}{\hat{\mathit{D}}}
\newcommand{\Lm}{\hat{\mathcal{L}}}
\newcommand{\Ein}{{\mathcal E}_{\rm in}}
\newcommand{\neffs}{N_{\rm eff}}%
\newcommand{\ld}{L_{\rm D,1}}

\newcommand{\deff}{d_{\rm eff}}

\newcommand{\lgvm}{L_{\rm GVM}}

\newcommand{\kpp}{\kz^{(2)}}
\newcommand{\tin}{T_{\rm in}}

\newcommand{\mic}{~{\rm \mu m}}

\include{00README.XXX}
\begin{document}


\twocolumn[ 
\title{Designing microstructured polymer optical fibers for cascaded
  quadratic soliton compression of femtosecond pulses}
\author{Morten Bache}\email{moba@fotonik.dtu.dk}
\address{DTU Fotonik, Department of Photonics Engineering,
  Technical University
of Denmark, DK-2800 Kgs. Lyngby, Denmark}%
\date{\today}
\begin{abstract}
  The dispersion of index-guiding microstructured polymer optical
  fibers is calculated for second-harmonic generation. The quadratic
  nonlinearity is assumed to come from poling of the polymer, which in
  this study is chosen to be the cyclic olefin copolymer Topas. We
  found a very large phase mismatch between the pump and the
  second-harmonic waves. Therefore the potential for cascaded
  quadratic second-harmonic generation is investigated in particular
  for soliton compression of fs pulses. We found that excitation of
  temporal solitons from cascaded quadratic nonlinearities requires an
  effective quadratic nonlinearity of 5 pm/V or more. This might
  be reduced if a polymer with a low Kerr nonlinear refractive index
  is used. We also found that the group-velocity mismatch could be
  minimized if the design parameters of the microstructured fiber are
  chosen so the relative hole size is large and the hole pitch is on
  the order of the pump wavelength. Almost all design-parameter
  combinations resulted in cascaded effects in the stationary regime,
  where efficient and clean soliton compression can be found. We
  therefore did not see any benefit from choosing a fiber design where
  the group-velocity mismatch was minimized. Instead numerical
  simulations showed excellent compression of $\lambda=800$ nm 120 fs
  pulses with nJ pulse energy to few-cycle duration using a standard
  endlessly single-mode design with a relative hole size of 0.4.
\end{abstract}

\maketitle 
]
\section{\label{sec:level1}Introduction} 

Microstructured optical fibers allows for an elaborate dispersion
control. Usually they are exploited to gain control over the
group-velocity dispersion (GVD) \cite{birks:1999} (see
\cite{laegsgaard:2007} for a recent literature overview over the
dispersion properties of microstructured optical fibers). Also the
group-velocity mismatch (GVM) for three-wave mixing processes can be
completely removed \cite{bache:2005a}, which is important in order to
realize second-harmonic generation (SHG) of ultra-short fs pulses.
However, the prize for having a small GVM is that the phase mismatch
becomes very large \cite{bache:2005a}, and efficient SHG would
therefore rely on quasi-phase matching (QPM) techniques.

The large phase-mismatch $\Delta \beta$ can instead be exploited for
cascaded $\chi^{(2)}:\chi^{(2)}$ SHG processes \cite{stegeman:1996}.
The second harmonic (SH) is within a coherence length $2\pi/|\Delta
\beta|$ generated and then back-converted to the fundamental wave
(FW).  This cascaded process generates a nonlinear phase shift
$\Phi_{\rm NL}$ on the FW \cite{desalvo:1992}, which can
become several units of $\pi$ by propagating in just a few cm's of
nonlinear material, and is conceptually equivalent to the nonlinear
phase shift observed through self-phase modulation (SPM) with cubic
nonlinearities \cite{menyuk:1994}.

The large nonlinear phase shift can be exploited to compress
ultra-short fs pulses \cite{liu:1999}. This is particularly
advantageous when the compression occurs inside the nonlinear material
due to the soliton effect \cite{ashihara:2002,moses:2006}; the
cascaded SHG process can generate a negative $\Phi_{\rm NL}$ resulting
in a negatively chirped FW, and if the material has normal GVD a
temporal soliton is generated. By launching a higher-order soliton,
the input pulse can be compressed by exploiting the initial pulse
narrowing. This cascaded quadratic soliton
compressor (CQSC) can in principle compress the FW pulse to single-cycle
duration, and is ultimately limited by higher-order dispersion
and competing Kerr nonlinear effects \cite{bache:2008}.

One requirement for efficient compression is that GVM effects are not
too strong: they tend to distort the compressed pulses through a
Raman-like effect \cite{ilday:2004,moses:2006,bache:2007a,bache:2008}.
In fact, when GVM dominates over the cascaded effects from the phase
mismatch the compression is \textit{nonstationary}, resulting in
inefficient compression and distorted pulses. Therefore the
possibility offered by microstructured optical fibers to control GVM
is very intriguing for the CQSC, because the compression can become
\textit{stationary}, which implies efficient compression and clean
pulses. The fiber geometry can also help overcoming the
problem of inhomogeneous compression in the transverse direction of
the beam found in a bulk geometry \cite{moses:2007}.\footnote{No comparison
should otherwise be made between the bulk and the fiber CQSC; the bulk
version works with high-energy fs pulses with $\mu$J-mJ energies,
while the fiber version works with low-energy fs pulses with pJ-nJ
energies.}

We have done a preliminary investigation of the potential in using
silica microstructured optical fibers for CQSC
\cite{bache:2007b,laegsgaard:2007}, where the quadratic nonlinearity
was assumed to come from thermal poling of the silica fiber
\cite{faccio:2001}. We surprisingly found that zero GVM was not as
such an advantage, since the compressed pulses became very distorted.
Another main result was that a very large quadratic nonlinearity
$\deff\sim 3-5$ pm/V was needed in order to generate suitably large
$\Phi_{\rm NL}$ \cite{laegsgaard:2007}.  This is because the chosen
fiber designs had very a large phase mismatch, and $\Phi_{\rm
  NL}\propto \deff^2/\Delta \beta$.  Realistically, thermal poling of
silica fibers can generate $\deff\sim 0.5$ pm/V
\cite{kazansky:private}. Therefore we suggested to lower the phase
mismatch using QPM techniques, and in this way a much lower $\deff\sim
1$ pm/V was sufficient \cite{bache:2007b}. However, periodic thermal
poling of silica fibers have until now proved inefficient, and has
resulted in a much lower $\deff\sim0.01$ pm/V than expected
\cite{Kazansky:1997}. We must therefore conclude that periodically poled
silica (microstructured) fibers cannot generate large enough
$\Phi_{\rm NL}$ to be interesting for CQSC.

A large $\deff$ would compensate for a large $\Delta \beta$. In this
paper, we therefore turn our attention to microstructured polymer
optical fibers (mPOFs). Poling of polymers can generate extremely
large quadratic nonlinearities (ranging from 1 pm/V to 100's of pm/V)
\cite{chang:2005}, while poling still has to be shown in a fiber
context.  Alternatively, the low mPOF drawing temperature (few hundred
degrees C) implies that nanomaterials with large quadratic nonlinear
responses (such as nanotubes \cite{guo:2005}) can be drawn into the
fiber core.  Also this solution promises a strong $\deff$. The
potential for using mPOFs for quadratic nonlinear optics is therefore
large.

Here we calculate the dispersion properties of an index-guiding mPOF
with three rings of air-holes in the cladding. The polymer material is
chosen to be the cyclic olefin copolymer Topas due to its broad
transparency window \cite{Khanarian:2001}. We point out that a problem
with using polymer as fiber material is that unlike silica the
material Kerr nonlinearity is very large, and it generates an
SPM-induced positive nonlinear phase shift which has to be overcome by
the negative cascaded nonlinear phase shift. We will show that for the
considered polymer Topas $\deff> 5$ pm/V is needed to do so (compared
to $\deff\sim 3-5$ pm/V in silica \cite{bache:2007b,laegsgaard:2007}).
Such a value could be achieved with polymers as fiber material. We
also show that the fiber designs with a dramatically reduced GVM are
multi-moded in the SH, and thus one risks to have cascaded nonlinear
conversion to several SH modes. Quite surprisingly we find that the
compression is always in the stationary regime, irrespective of the
choice of fiber-design parameters. This is very positive since
efficient and clean compression can be obtained. On the other hand,
there is no longer any motivation for reducing GVM in order to enter
the stationary regime. Thus, we conclude that there are very few
benefits of reducing GVM through fiber design. This statement is
underlined by performing numerical simulations of the propagation
equations for an mPOF having $\deff=10$ pm/V: an endlessly
single-moded mPOF design dominated mainly by material dispersion turns
out to give clean and efficient compression of fs pulses, and obtaining
compressed pulses with durations close to single-cycle duration is
possible. So if the potentially large $\deff$ of polymer material can
be exploited, then excellent pulse compression can be obtained in
mPOFs.

\section{Nonlocal theory}
\label{sec:Nonlocal-theory}

The scope of this theoretical part is to point out the important
parameters when choosing the proper fiber design. The main hypothesis
is that an mPOF can be used as a CQSC. The challenges with creating a
quadratic nonlinearity in the fiber aside, there are other obstacles
that must be faced. Understanding these issues can be greatly enhanced
by realizing that in the cascading regime, the coupled FW and SH
propagation equations 
reduce to a nonlinear Schr{\"o}dinger equation
(NLSE) for the FW \cite{menyuk:1994}.
The action of the cascaded SHG can be modeled as a Kerr-like
nonlinearity with an equivalent nonlinear refractive index $n_{\rm
  SHG}^I\propto\deff^2/\Delta \beta$.  Importantly, when $\Delta
\beta=\beta_2-2\beta_1>0$ (which is usually the case) $n_{\rm SHG}^I$
is self-defocusing of nature: it generates a negative nonlinear phase
shift $\Phi_{\rm NL}<0$. The material Kerr nonlinear refractive
index\footnote{The Kerr nonlinear refractive index is usually denoted
  $n_2$, but we reserve this subscript to the SH.} $n_{\rm Kerr}^I$
is instead usually self-focusing, and therefore counteracts the
effects of the cascaded nonlinearities. Achieving $n_{\rm
  SHG}^I>n_{\rm Kerr}^I$ is crucial to obtain a large negative
nonlinear phase shift. An important point here is that because the
cascaded SHG induces a self-defocusing Kerr-like nonlinearity,
temporal solitons exists in presence of normal GVD. In contrast, the
usual cubic temporal solitons observed for instance in telecom fibers
require anomalous GVD due to the self-focusing Kerr nonlinearity of
silica \cite{mollenauer:1980}.

GVM is also playing a decisive role. As we recently showed, the pulse
compression is clean and efficient in the stationary regime, where the
phase mismatch dominates over GVM effects
\cite{bache:2007a,bache:2008}. These results were obtained by deriving
a more general NLSE, where dispersion including GVM imposes a
Raman-like nonlocal (delayed) temporal cubic nonlinearity. Contrary to
earlier reports \cite{liu:1999,ilday:2004} we showed that the regime
where efficient compression takes place is independent on the input
pulse duration.

In the remainder of this section we briefly show how the nonlocal NLSE
for the FW is derived, and discuss the consequences for soliton pulse
compression. The equation is derived in dimensionless form based on
the SHG propagation equations derived in
App.~\ref{sec:Gener-prop-equat}.

The nonlocal theory was derived in Refs.
\cite{bache:2007a,bache:2008} (for a general review on nonlocal
effects, see \cite{krolikowski:2004}). Essentially in the cascading limit
(large phase mismatch), the SH becomes slaved to the FW. The
normalized FW $U_1$ can then on dimensionless form be modeled by the
following \textit{nonlocal} nonlinear Schr{\"o}dinger equation (NLSE)
\cite{bache:2007a}
\begin{align}
  \label{eq:fh-shg-nlse-nonlocal}
\left[i\frac{\partial}{\partial\xi}-
  \frac{{\rm sgn}(\kpp_1)}{2}\frac{\partial^2}{\partial\tau^2}\right]U_1
  +\nkerr^2U_1|U_1|^2 
\\ 
\nonumber
=
\nshg^2U_1^*\int_{-\infty}^{\infty} {\rm d}s
  R_\pm(s)U_1^2(\xi,\tau-s).
\end{align}
where Kerr cross-phase modulation (XPM) effects have been neglected,
and for simplicity self-steepening and higher-order dispersion are not
considered. On the left-hand side an ordinary NLSE appears with an
SPM-term from self-focusing material Kerr nonlinearities. On the
right-hand side the effects of the cascaded SHG appear: it is also
SPM-like in nature, but it turns out to be controlled by a temporal
nonlocal response. The dimensionless temporal nonlocal response
function $R_\pm(\tau)$ appears in two distinct ways: in the
\textit{stationary regime} it is given by $R_+$, and $|R_+|\propto
e^{-|\tau|/\tau_b}$ is localized having a characteristic width
$\tau_b$ of a few fs. This happens when the phase mismatch is
dominating over GVM effects, or more precisely when $\Delta\kz>\Delta
\kz_{\rm sr}$, with
\begin{equation}\label{eq:stationary}
\Delta \kz_{\rm sr}=\frac{d_{12}^2}{2\kpp_2}=-\frac{\pi
  d_{12}^2}{{\mathcal D}_2\lambda_2^2}. 
\end{equation}
and ${\mathcal D}_2=-2\pi\kpp_2/\lambda_2^2$ is the fiber GVD
parameter.  In the stationary regime clean and efficient compression
can be obtained \cite{moses:2006,bache:2007a,bache:2007,bache:2008}.
Instead in the \textit{nonstationary regime} ($\Delta\kz<\Delta
\kz_{\rm sr}$) GVM effects dominate, and the compression becomes
distorted and inefficient. In this case the nonlocal response function
is given by $R_-$, and $|R_-|\propto\sin(|\tau|/\tau_b)$ is
oscillatory and never decays.

The fiber designs presented here all turn out to be in the stationary
regime. In this case, and when the nonlocal response function can be
assumed quasi-instantaneous, Eq.~(\ref{eq:fh-shg-nlse-nonlocal}) can
be written as \cite{ilday:2004,moses:2006,bache:2007a}
\begin{multline}
  \label{eq:fh-shg-nlse-weakly-nonlocal}
  \left[i\frac{\partial}{\partial\xi}-
  \frac{{\rm sgn}(\kpp_1)}{2}\frac{\partial^2}{\partial\tau^2}\right]U_1 \\ 
  -[{\rm sgn}(\Delta \kz)\nshg^2-\nkerr^2] U_1|U_1|^2 
\\ 
=-i\nshg^2s_a\tau_{R,\rm SHG}|U_1|^2\frac{\partial
  U_1}{\partial \tau}
\end{multline}
The soliton order for the Kerr fiber nonlinearity is
\cite{agrawal:1989}
\begin{align}
  \label{eq:Nkerr-bulk}
  \nkerr^2=\frac{\ld}{L_{\rm Kerr}}
\end{align}
where $\ld=\tin^2/|\kpp_1|$ is the characteristic GVD length of the
FW, and $L_{\rm Kerr}=(\gamma_{\rm Kerr}\Pin)^{-1}$ is the
characteristic Kerr nonlinear length. $\Pin= \varepsilon_0
\neff{1}ca_1\Ein^2/2$ is the input peak power, related to the peak
input electric field $\Ein$. The Kerr nonlinear coefficient is
\begin{align}\label{eq:gamma-kerr}
\gamma_{\rm Kerr}=n_{\rm Kerr}^I\frac{\omega_1}{c\Aeff{1}}
\end{align}
where $\Aeff{1}$ is the effective Kerr FW mode overlap area
\begin{align}\label{eq:FW-kerr-ovl}
\Aeff{1}= \frac{(\int d\x |F_1(\x)|^2)^2} {\int_P
  d\x    |F_1(\x)|^4} 
\end{align}
Note that $\Aeff{1}\equiv f_{11}^{-1}$ from
Eq.~(\ref{eq:Kerr-ovl-norm}), and that as explained in
App.~\ref{sec:Gener-prop-equat} the subscript 'P' indicates that
integration must be done only over the part of the fiber mode sitting
in the polymer part of the mPOF. We also encounter the well-known Kerr
nonlinear refractive index $n_{\rm Kerr}^I$: the refractive index
change due to the Kerr self-focusing effect is defined as $\Delta
n=n_{\rm Kerr}^I I$, where $I$ is the intensity of the beam. Although
it is not known for the Topas polymer, we estimate (based on similar
polymer types) that it is in the range $n_{\rm Kerr}^I\sim 10-15 \cdot
10^{-20}~{\rm m^2/W}$, which is an order of magnitude larger than for
fused silica.

In a similar way we have introduced a soliton order from the cascaded
SHG process
\begin{align}  \label{eq:Nshg-bulk}
  \nshg^2=\frac{\ld}{L_{\rm SHG}}
\end{align}
where $L_{\rm SHG}=(\gamma_{\rm SHG}\Pin)^{-1}$. Here the nonlinear
coefficient is
defined as for the Kerr case
\begin{align}
\gamma_{\rm SHG}=|n_{\rm SHG}^I|\frac{\omega_1}{c \Aovl }
\end{align}
only it contains the SHG overlap area
$\Aovl$
\begin{align}\label{eq:A-ovl-shg}
  \Aovl=\frac{a_1^2a_2}{|\int_P  
  d\x [F_1^*(\x)]^2F_2(\x)|^2}
\end{align}
where $a_j$ are the fiber mode areas~(\ref{eq:modearea}). The
``effective'' nonlinear refractive index from the cascaded process
\cite{desalvo:1992}
\begin{align}
  n_{\rm
  SHG}^I=-\frac{4\pi\deff^2}{c\varepsilon_0\lambda\neff{1}^2\neff{2}\Delta\kz} 
\end{align}
Note the negative sign in front of $ n_{\rm SHG}^I$; since the cases we will
discuss always have $\Delta \kz>0$ the cascaded nonlinearity is therefore
self-defocusing. Thus, in order to generate a negative nonlinear phase
shift we must have $\gamma_{\rm SHG}>\gamma_{\rm Kerr}$. We can
express this through an effective soliton order
\begin{align}
  \neffs^2=\nshg^2-\nkerr^2=\Pin\ld\left(\gamma_{\rm SHG}-\gamma_{\rm Kerr}\right)
\end{align}
Using this in Eq.~(\ref{eq:fh-shg-nlse-weakly-nonlocal}) we then get
an NLSE with a self-defocusing SPM term:
\begin{multline}
  \label{eq:fh-shg-nlse-weakly-nonlocal1}
  \left[i\frac{\partial}{\partial\xi}-
  \frac{1}{2}\frac{\partial^2}{\partial\tau^2}\right]U_1  
  -\neffs^2 U_1|U_1|^2 
\\ 
=-i\nshg^2s_a\tau_{R,\rm SHG}|U_1|^2\frac{\partial
  U_1}{\partial \tau}
\end{multline}
Self-defocusing solitons with strength $\neffs^2$ can then be excited
if the FW GVD is normal, i.e. if $\kpp_1>0$, which has been taken into
account in Eq.~(\ref{eq:fh-shg-nlse-weakly-nonlocal1}).

So the LHS of Eq.~(\ref{eq:fh-shg-nlse-weakly-nonlocal1}) is now a
self-defocusing NLSE supporting solitons if $\neffs\geq 1$. The RHS
contains the Raman-like perturbation due to GVM. It becomes important
for large SHG soliton orders $\nshg$ \cite{bache:2008} and when the
compressed pulse duration is on the order of $T_{R, \rm SHG}=\tau_{R,
  \rm SHG}\tin$, where the characteristic dimensionless time scale of
the Raman-like perturbation is
\begin{align}\label{eq:trshg}
  \tau_{R, \rm SHG}\equiv \frac{2|d_{12}|}{\Delta \kz\tin}
\end{align}
Typically $T_{R, \rm SHG}$ is on the order of 1-5 fs, but can become
very large in the nonstationary regime \cite{bache:2008}. Finally,
$s_a={\rm sgn}(d_{12}\kpp_2)$: the Raman-like effect will depending on
this sign give either a red-shifting ($s_a=+1$) or blue-shifting
($s_a=-1$) of the FW spectrum. In the cases we present here $s_a=-1$.

\section{Numerical results}
\label{sec:Numerical-results}

In this section the transverse fiber modes and their dispersion
properties are calculated for various mPOF designs (varying the
air-hole diameter $d$ and the hole pitch $\Lambda$). Then we show
compression examples for selected fiber designs. These numerical
simulations were done using the full coupled propagation
equations~(\ref{eq:shg-bulk-U}), which include Kerr XPM effects,
steepening terms, and are valid down to single-cycle resolution using
the slowly-evolving wave approximation
\cite{brabec:1997,moses:2006b,bache:2007}.

\subsection{The microstructured fiber}
\label{sec:transverse-fiber-mod}

\begin{figure}[tb]
  \begin{center}
\includegraphics[width=4cm]{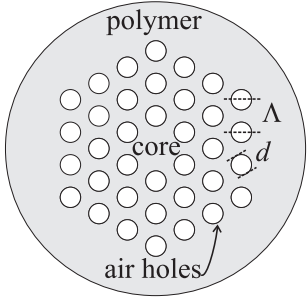}
  \end{center}
\caption{\label{fig:mPOF} The microstructured fiber design considered
  here. }
\end{figure}

We consider an index-guiding mPOF with a triangular air-hole pattern
in the cladding, see Fig.~\ref{fig:mPOF}.  The mPOF design parameters
are the pitch $\Lambda$ between the air holes and the relative air
hole size $D=d/\Lambda$, where $d$ is the physical air-hole diameter.
We assume that the core has a quadratic nonlinearity from thermal
poling of the polymer after the fiber drawing, or from including
nonlinear nanomaterial in the fiber preform.

We consider an mPOF made from the cyclic olefin copolymer Topas
\cite{Khanarian:2001}. The advantage of using Topas compared to, e.g.,
poly-(methyl methacrylate) (PMMA), is that Topas has a larger
transparency window: it is transparent from 300-1700 nm, interrupted
by two absorption peaks around 1200 nm and 1450 nm
\cite{Khanarian:2001}, see App.~\ref{sec:Topas} for more details.

\subsection{Fiber design, dispersion and compression}
\label{sec:Fiber-design-disp}

We will here search for fiber design parameters, where the requirement
is an optimal dispersion profile for cascaded quadratic soliton
compression. Details about the fiber mode calculations and the
dispersion calculations are found in App.~\ref{sec:Calc-fiber-modes}.

Two types of designs are discussed: one is a realistic mPOF with a
large air-hole pitch of around $\Lambda=7\mic$ that can easily be
drawn, while another takes $\Lambda$ small (comparable to the FW
wavelength) as to significantly alter the dispersion parameters. Such
an mPOF would probably need to be done by tapering an mPOF with a
larger pitch. In the calculations the Sellmeier equation from
App.~\ref{sec:Topas} with $T=25^\circ$C was used.

\begin{figure}[tb]
  \begin{center}
    \includegraphics[width=8cm]{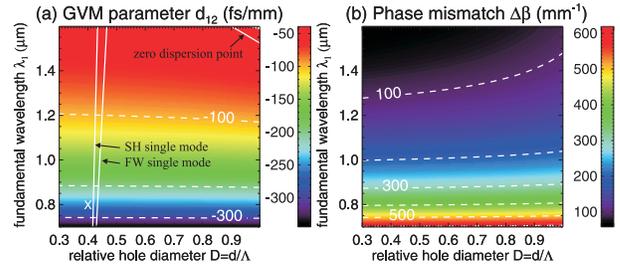}
  \end{center}
\caption{\label{fig:dvg_large} The GVM parameter $d_{12}$ and phase
  mismatch $\Delta \beta$ for an mPOF with $\Lambda=7.0\mic$. In (a)
  the lines for single-mode operation are indicated for the FW and the
  SH modes. The FW zero dispersion point is also indicated; below this
  line the FW GVD is normal. Finally the 'X' indicates the chosen fiber
  design for single-mode operation at $\lambda_1=800$ nm. }
\end{figure}

\begin{figure}[tb]
  \begin{center}
\includegraphics[width=7.5cm]{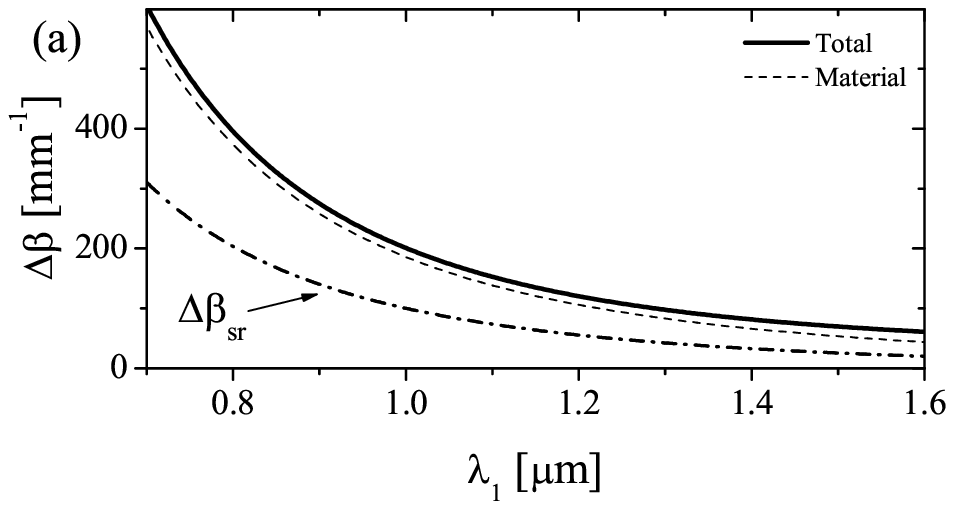}
\includegraphics[width=7.5cm]{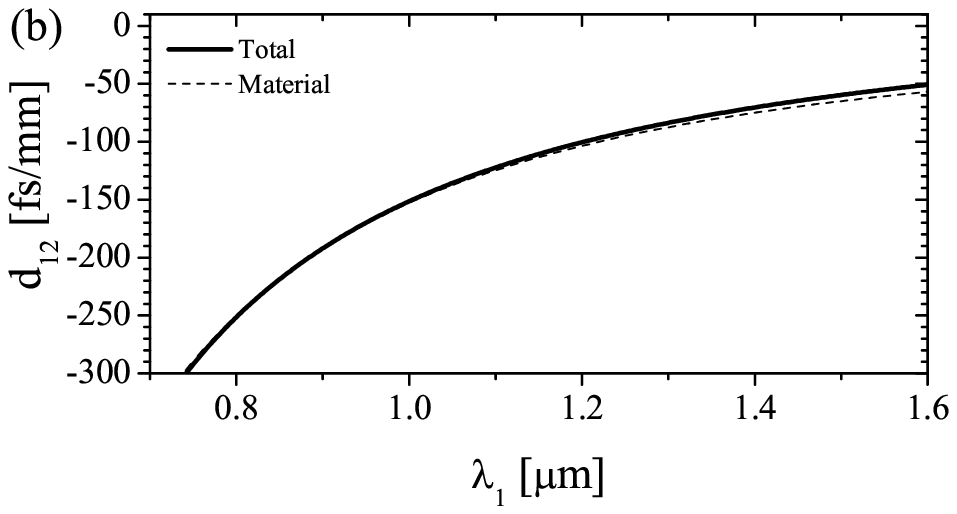}
\includegraphics[width=7.5cm]{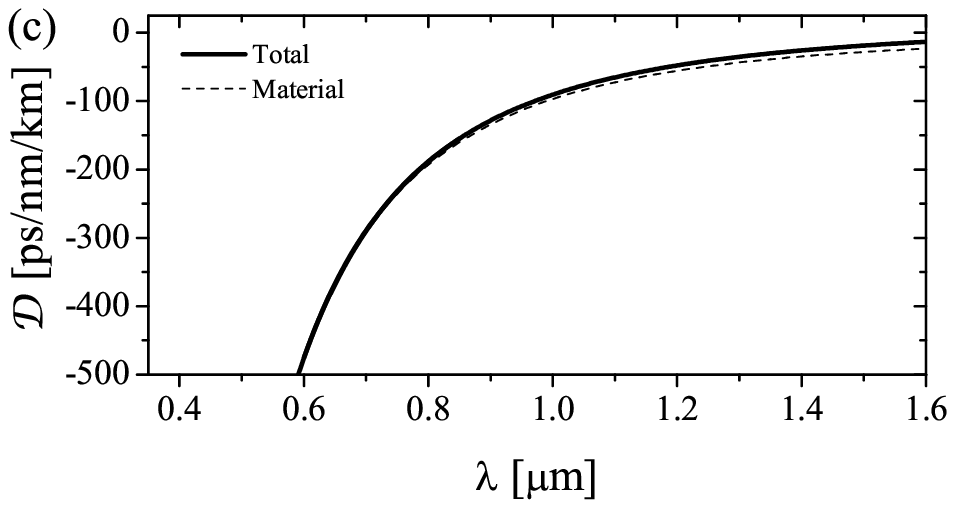}
\includegraphics[width=7.6cm]{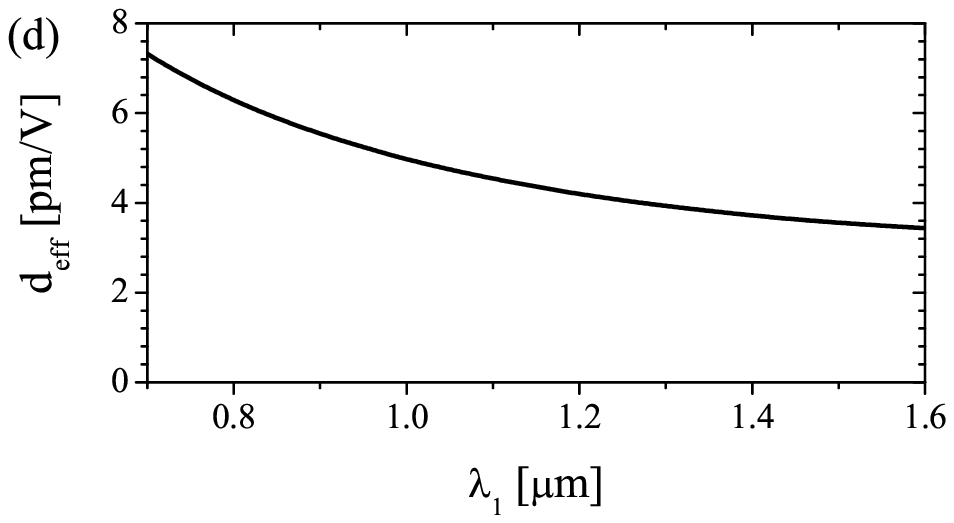}
  \end{center}
\caption{\label{fig:disp_large} The dispersion for 
  $\Lambda=7.0\mic$ and $D=d/\Lambda=0.4$. (a) the
  phase mismatch $\Delta \beta=\beta_2-2\beta_1$ and (b) GVM
  $d_{12}=\beta_1^{(1)}-\beta_2^{(2)}$ vs. $\lambda_1$. (c)
  the GVD parameter ${\mathcal D}=-2\pi c \beta^{(2)}/\lambda^2$
  vs. $\lambda$. The material dispersion
  is also shown. (d) shows the effective
  nonlinearity required to achieve $\gamma_{\rm SHG}>\gamma_{\rm
    Kerr}$ for $n_{\rm Kerr}^I=15\cdot 10^{-20}~{\rm m^2/W}$.}
\end{figure}

It is fairly standard to draw mPOFs with a air-hole pith of
$\Lambda>5\mic$. With this as a starting point we show in
Fig.~\ref{fig:dvg_large} the GVM and phase mismatch for a Topas mPOF
with $\Lambda=7.0\mic$. We notice that the choice of $D$ does not
affect the dispersion parameters much. This is because the size of the
waveguide [the core diameter is $d_{\rm core}=\Lambda(2-D)$] is
significantly larger than the wavelength of the guided light, and so
material dispersion is dominating (see also Fig.~\ref{fig:disp_large}). 
The microstructured cladding therefore alters the dispersion only very
little, and we cannot get tailored GVM or GVD properties.

Since Topas has a primary transparency window between 290-1210 nm, we
focus now on SHG with $\lambda_1=800$ nm. We would like the fiber to
be single mode both at the FW and at the SH wavelength, because then
phase matching can only occur between these modes and we do not have
problems with the typical higher-order mode interaction usually
observed in wave-guided SHG. The criterion for single-mode operation in
the particular microstructured fiber, we investigate, is \cite{kuhlmey:2002}
$ \lambda/\Lambda=2.8(D-0.406)^{0.89} $. 
This criterion gives the lines shown in Fig.~\ref{fig:dvg_large}(a)
for the FW (where $\lambda=\lambda_1$) and the SH (where
$\lambda=\lambda_2=\lambda_1/2$). We therefore choose a design with
$D=0.4$, which is indicated with an 'X'; this design is actually
endlessly single-moded \cite{Birks:1997}. 

For the chosen design, we show in Fig.~\ref{fig:disp_large} the
dispersion parameters as function of wavelength. 
In (a) the phase mismatch $\Delta \beta$ is shown. It is always above
the boundary to the stationary regime as given by
Eq.~(\ref{eq:stationary}) $\Delta \beta_{\rm
  sr}=d_{12}^2/2\beta_2^{(2)}$. Thus, the chosen fiber design for all
the shown wavelengths is in the stationary regime for soliton
compression, which is very good news for the possibility of clean and
efficient compression. This might seem surprising given the large GVM
seem in Fig.~\ref{fig:disp_large}(b), but it is a consequence of having a
very large phase mismatch and the very large GVD at the SH wavelength,
see Fig.~\ref{fig:disp_large}(c).  However, the large phase mismatch
unfortunately also implies a small cascaded nonlinear parameter
$\gamma_{\rm SHG}$. In order to achieve $\gamma_{\rm SHG}>\gamma_{\rm
  Kerr}$, as required for solitons to exist, the effective quadratic
nonlinearity $\deff$ must be quite high, around 6 pm/V for
$\lambda_1=800$ nm as shown in Fig.~\ref{fig:disp_large}(d).

\begin{figure}[tb]
  \begin{center}
\includegraphics[width=8.5cm]{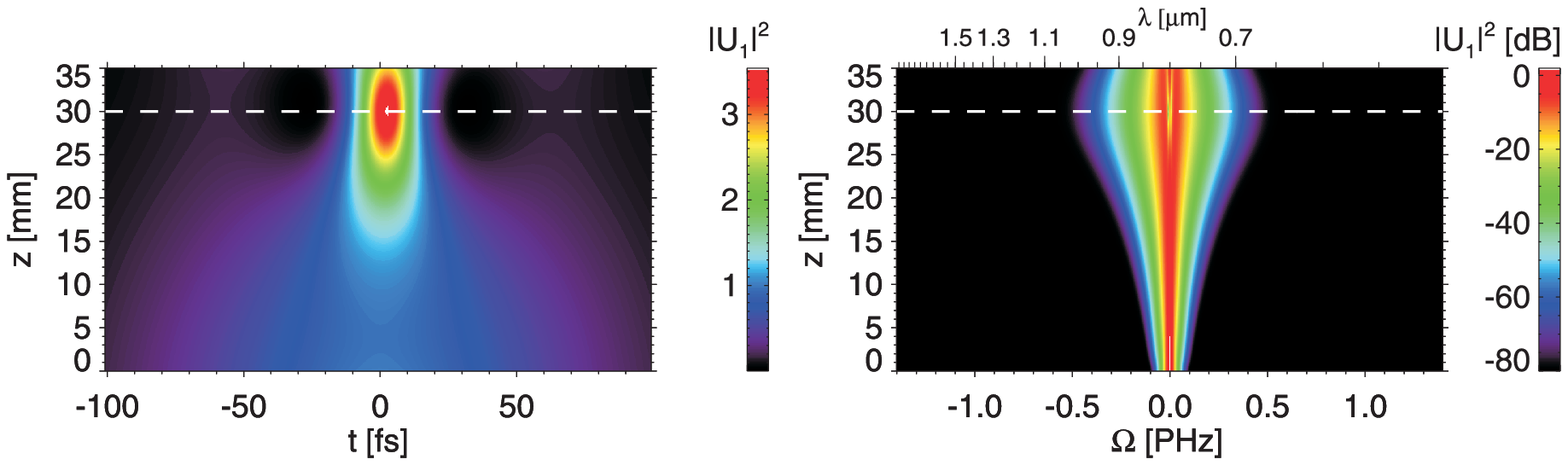}\\
\includegraphics[width=8.5cm]{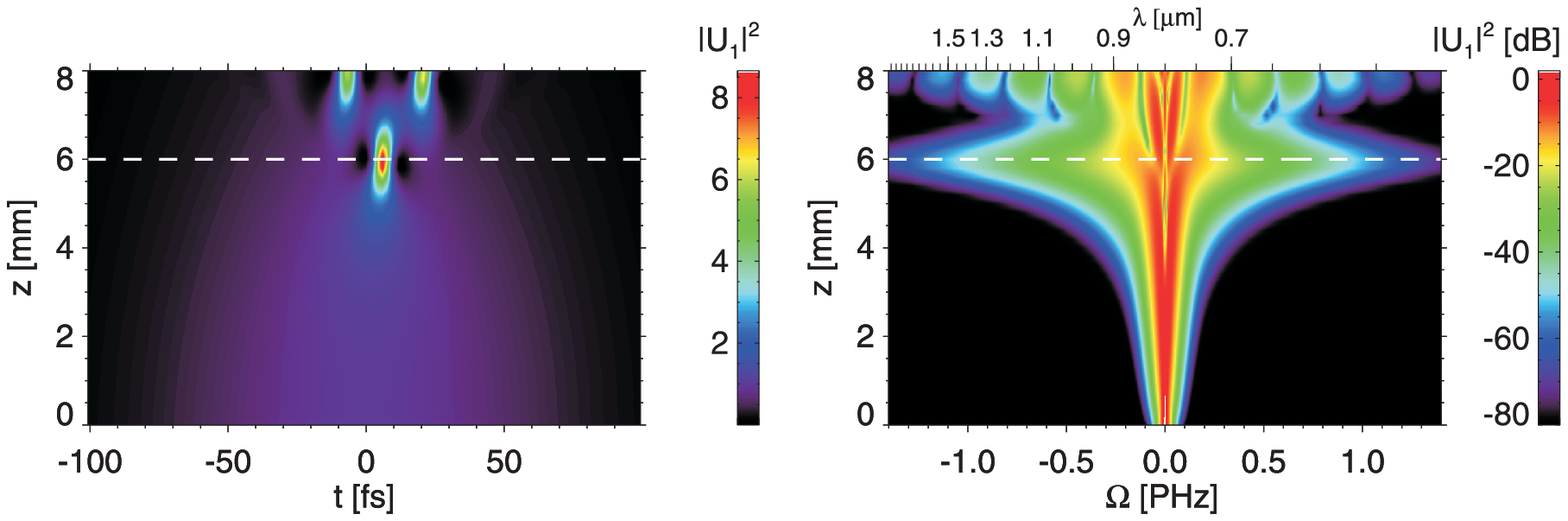}\\
\includegraphics[width=8.5cm]{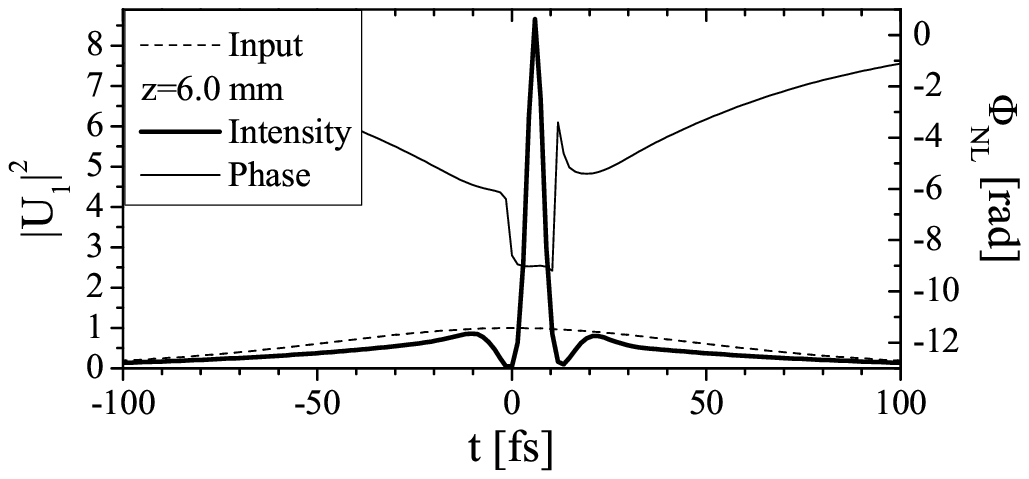}    
  \end{center}
\caption{\label{fig:compr_large-2D} Numerical simulation of soliton
  compression in an mPOF with the design parameters of
  Fig.~\ref{fig:disp_large} and $\deff=10$ pm/V. Top left plot shows
  the FW compressing to 15 fs FWHM after 30 mm of propagation (at the
  dashed line) starting from 120 fs. Top right shows the spectral
  broadening and development of SPM-induced sidebands of the FW. The
  pulse parameters were $\neffs=3$, $\Pin=4.6$ kW, and the pulse
  energy was 0.6 nJ. The middle row shows a simulation for a higher soliton
  order ($\neffs=9$, $\Pin=41.8$ kW, and a pulse energy of 5.7 nJ),
  which compresses to 4.4 fs after 6 mm of propagation. A cut at $z=6$
  mm is shown in the bottom plot. Up to 10th order
  dispersion was included ($m_d=10$). $2^{13}$ temporal points and
  $> 15$ $z$-steps per coherence length were used.}
\end{figure}

In Fig.~\ref{fig:compr_large-2D} we show numerical simulations of
soliton compression using the chosen design. For the Kerr nonlinear
refractive index of the polymer Topas we use $n_{\rm Kerr}^I=15\cdot
10^{-20}~{\rm m^2/W}$, which is a large but realistic value. In order
to outbalance this Kerr self-focusing nonlinearity the quadratic
nonlinearity must be $\deff>6$ pm/V, see Fig.~\ref{fig:disp_large}(d),
and $\deff=10.0$ pm/V was chosen. The compression is shown for both a
low and a high soliton order, and both are very clean for several
reasons: firstly, the compression occurs in the stationary regime, and
secondly the Raman-like perturbation is very small, $T_{R,\rm
  SHG}=1.3$ fs. Lastly, looking at the FW frequency spectra there are
no dispersive waves. These tend to induce trailing oscillations on the
compressed pulse \cite{bache:2008}, but we checked that for this fiber
design they are not phase matched in the transparent region of
Topas. We should also mention that we have neglected any cubic Raman
effects of the material due to the lack of material knowledge in this
respect, and if significant Raman effects are present as for silica
fibers then this would have an impact on the results.

\begin{figure}[tb]
  \begin{center}
\includegraphics[width=8.5cm]{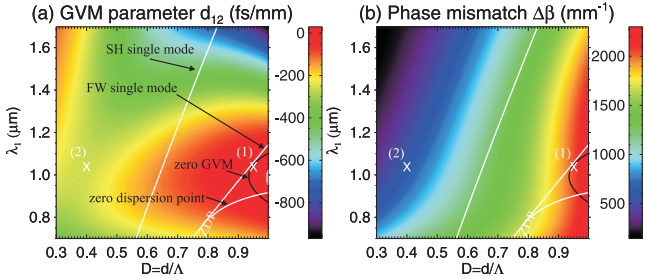}    
  \end{center}
\caption{\label{fig:dvg_small} As Fig.~\ref{fig:dvg_large} but with
  $\Lambda=0.65\mic$. A zero GVM curve is indicated with black as well
  as the FW zero dispersion point (ZDP). The two 'X'es indicates the
  chosen fiber designs at $\lambda_1=1040$ nm. }
\end{figure}

So a standard mPOF, where the fiber simply serves as a waveguide but
otherwise has little influence on the dispersion, can give decent
compression (provided $\deff$ is high enough). What benefits can we
get with a reduced GVM? 
To get a detailed understanding of this, let us
investigate a possible fiber design having very small GVM.

In order to significantly alter the dispersion, the mPOF must have a
lower hole pitch. As an extreme example Fig.~\ref{fig:dvg_small} shows
the calculated fiber properties for $\Lambda=0.65\mic$ \footnote{This
  value is unrealistic but serves the purpose of
  discussing the conditions under which zero or low GVM can be
  achieved.}: the GVM can be very small or even zero along the black
curve.  Unfortunately for $\lambda_1=800$ nm this low-GVM area is in
an anomalous dispersion region, where cascaded solitons do not exist,
so let us instead work at $\lambda_1=1040$ nm, which
would be relevant for compressing the output of Yb:doped fiber lasers.

\begin{figure}[htb]
  \begin{center}
\includegraphics[width=8cm]{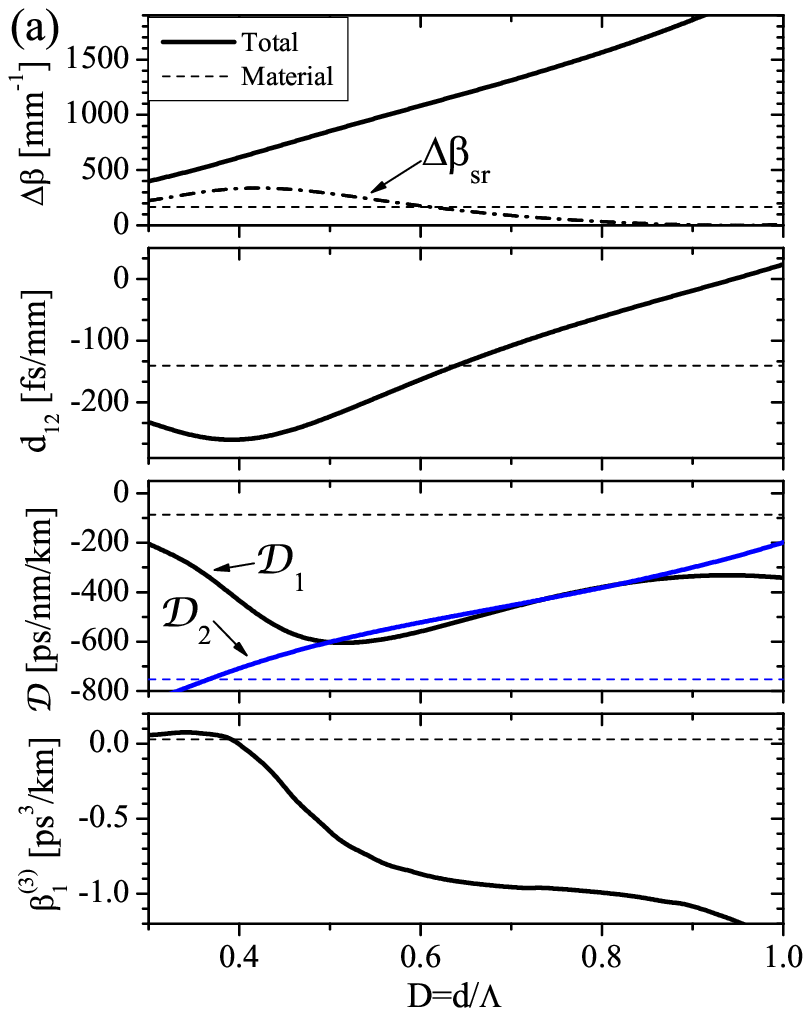}\\
\includegraphics[width=8cm]{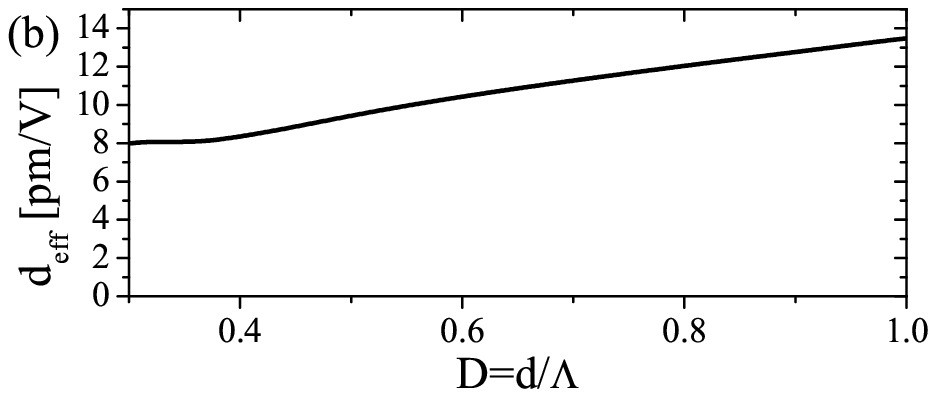}
  \end{center}
\caption{\label{fig:disp_small} (a) The dispersion as function of the
  relative hole size $D$ as calculated for
  $\lambda_1=1040$ nm and $\Lambda=0.65\mic$. (b) The the effective
  nonlinearity required to achieve $\gamma_{\rm SHG}>\gamma_{\rm
    Kerr}$ assuming $n_{\rm Kerr}^I=15\cdot 10^{-20}~{\rm m^2/W}$.}
\end{figure}

The dispersion for $\lambda_1=1040$ nm of this mPOF with
$\Lambda=0.65\mic$ is shown in Fig.~\ref{fig:disp_small}(a) as
function of $D$. Clearly the dispersion changes dramatically as $D$ is
changed.  The GVM goes from a normal value ($d_{12}\sim -200$ fs/mm)
for low $D$ to zero at high $D$. Nonetheless, we remain in the
stationary regime even for low $D$-values since $\Delta
\kz>\Delta\kz_{\rm sr}$ always.  This is because both $\Delta \beta$
and the SH GVD ${\mathcal D}_2$ are large.  Notice also that the FW
third order dispersion $\kz_1^{(3)}$ changes dramatically: it is very
large for high $D$ and close to zero for low $D$. Finally,
Fig.~\ref{fig:disp_small}(b) shows the critical $\deff$ required in
order to overcome the material Kerr self-focusing effects. It is
around 2-3 times higher than in Fig.~\ref{fig:disp_large}, which is a
consequence of a large $\Delta \beta$.

Zero GVM can therefore be achieved choosing $D=0.95$ [at the cross
marked (1) in Fig.~\ref{fig:dvg_small}]. This choice leaves the SH
multi-moded, and is therefore not suitable for compression: there
would be different modes with different degrees of phase mismatch
interacting with the FW, and gaining control over the compression
would be difficult. If we want the SH to be single moded, we should
choose a design below $D=0.6$ (see Fig.~\ref{fig:dvg_small}), and the
design marked with 'X' (2) in Fig.~\ref{fig:dvg_small} has the
advantage of being endlessly single moded. In this regime the GVM is
not too different from material dispersion.

So the penalty of choosing a small-pitched fiber in order to reduce
the GVM seems to be too high. The main motivation behind controlling
GVM through the fiber microstructure was to be in the stationary
regime.  However, almost all mPOF designs we have investigated have
been in the stationary regime thanks to the large phase mismatch.
Another motivation was to reduce the characteristic time of the
Raman-like response $T_{R,\rm SHG}=|d_{12}|/2\Delta \kz$ as to get
cleaner compressed pulses, but $T_{R,\rm SHG}$ is low even when GVM is
not reduced again thanks to the large phase mismatch. What is more,
choosing a zero GVM design surprisingly has been shown not to benefit
compression: In our previous study \cite{bache:2007b} we performed
simulations with just the lowest SH transverse mode (neglecting the
higher-order modes), and the fiber designs with zero GVM were giving
quite distorted compressed pulses.  This can be partially explained by
the fact that higher-order dispersion is significant when the relative
hole size is large; observe for instance how large the third-order
dispersion is in Fig.~\ref{fig:disp_small} for $D$ close to unity:
this value is several order of magnitude stronger than the material
third-order
dispersion. 

\begin{figure}[tb]
  \begin{center}
\includegraphics[width=8cm]{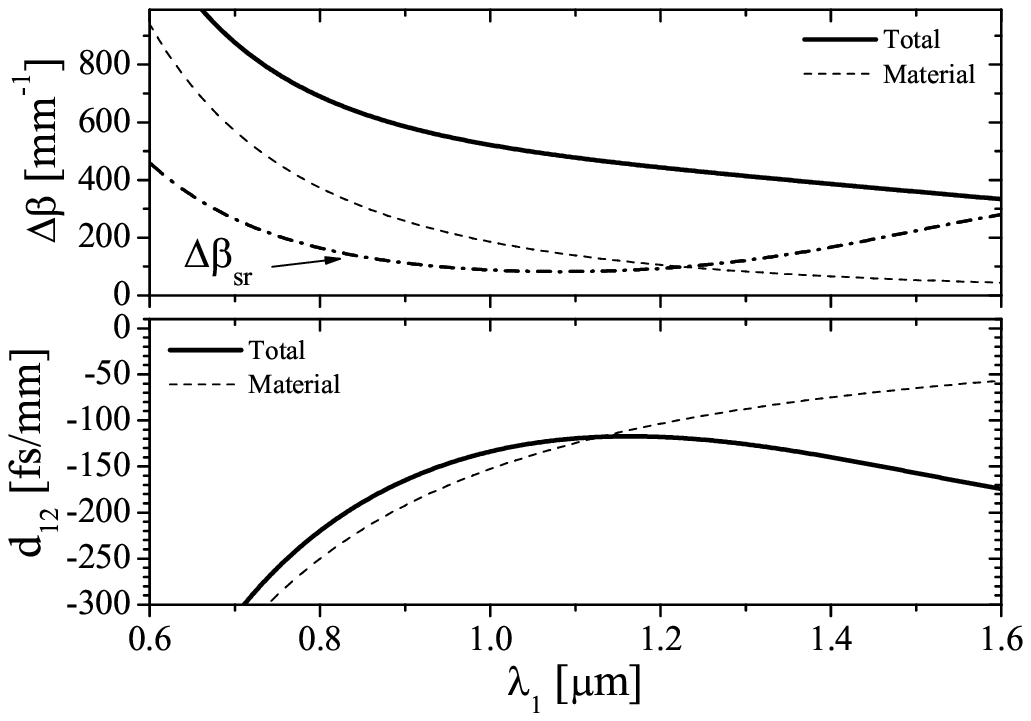}\\
\includegraphics[width=8cm]{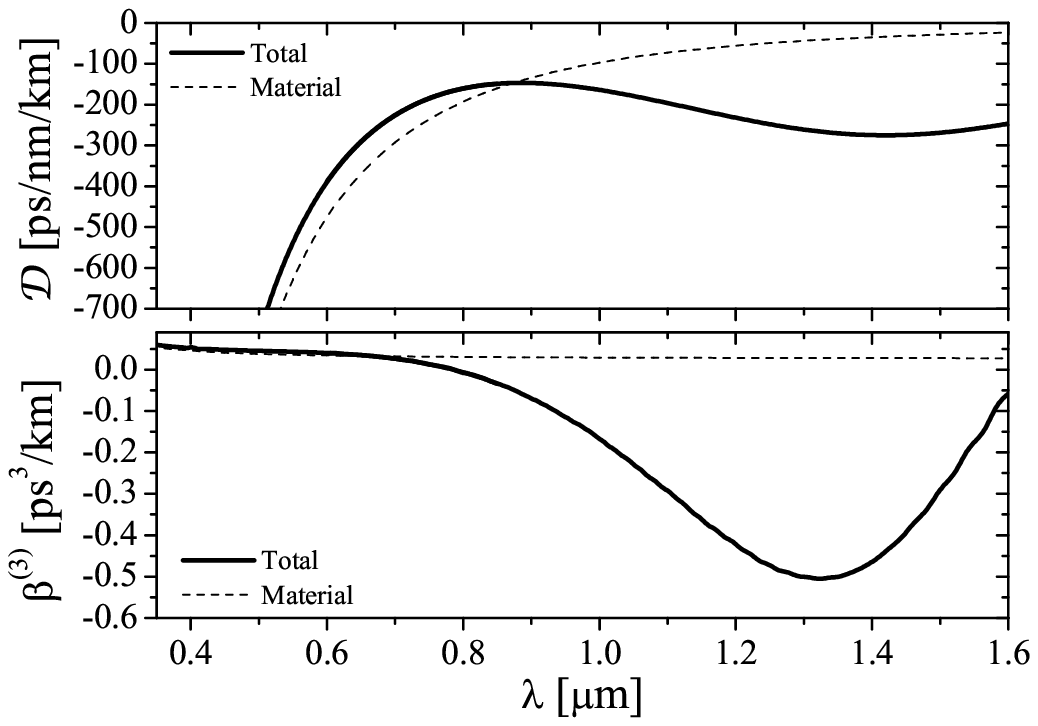}\\
\includegraphics[width=8cm]{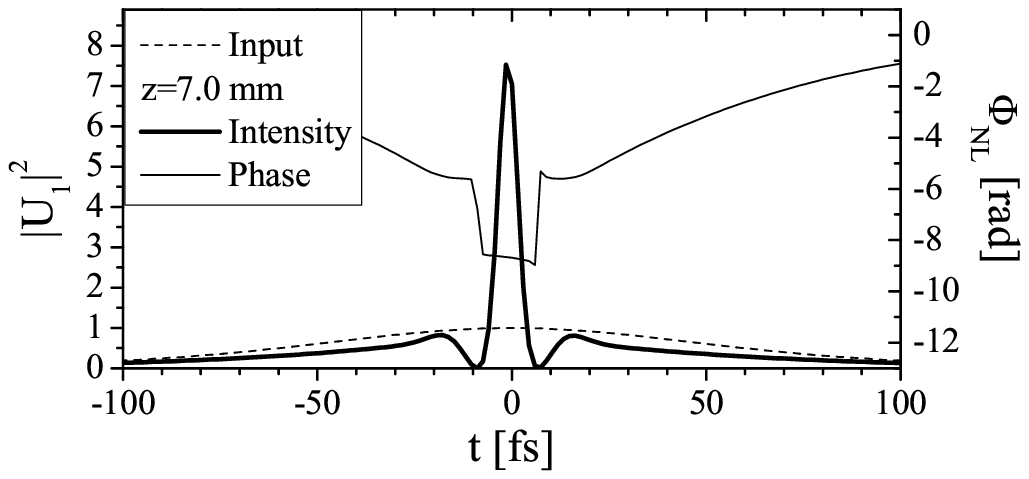}    
  \end{center}
\caption{\label{fig:disp_med} The dispersion calculated for
  $D=0.4$ nm and $\Lambda=1.0\mic$. The lower plot shows the results
  of a numerical simulation, where a $\lambda_1=800$ nm $\tin=120$ fs
  FWHM input pulse is compressed to 5.5 fs. The pulse cut is shown
  after $z=7.0$ mm of propagation. The fiber had $\deff=10$ pm/V as in
  Fig.~\ref{fig:compr_large-2D}. The input pulse had $\neffs=9$,
  $\Pin=3.1$ kW, and pulse energy 0.42 nJ. }
\end{figure}

An advantage of small-pitched fibers is instead that
nonlinear effects occur with a much smaller power. As an example,
choosing $\Lambda=1.0\mic$ and $D=0.4$ gives an endlessly single-mode
mPOF, where GVM and GVD are similar to the material dispersion (see
the calculated dispersion parameters in Fig.~\ref{fig:disp_med}).
This situation is therefore similar to the large-pitch case discussed
previously (where $\Lambda=7.0\mic$). However, due to the much smaller
core diameter ($d=1.6\mic$ compared to $d=11.2\mic$ in
Fig.~\ref{fig:compr_large-2D}) and mode areas ($a_1=0.43~\mic^2$
compared to $a_1=21.2~\mic^2$ in Fig.~\ref{fig:compr_large-2D}) a
compression similar to the one shown in Fig.~\ref{fig:compr_large-2D}
can be done with 10 times lower pulse energy and power: one can
compare the compressed pulse shown in Fig.~\ref{fig:disp_med} for
$\Lambda=1.0\mic$ with the one for $\Lambda=7.0\mic$ shown in
Fig.~\ref{fig:compr_large-2D}. Both have the same soliton order, and
achieve essentially the same level of compression, but the pump power
required is $\Pin=3.1$ kW when $\Lambda=1.0\mic$ and $\Pin=41.8$ kW
when $\Lambda=7.0\mic$.

\section{Conclusions}
\label{sec:Conclusions}

Our recent theory for cascaded quadratic soliton compression
\cite{bache:2007a,bache:2007,bache:2008} has shed new light on the
role of the dispersion in the compression of fs near-IR pulses with
cascaded SHG. But what can be gained by having the control over the
GVM that microstructured fibers offer \cite{bache:2005a}? An initial
study on silica microstructured fibers illustrated that zero GVM is
not a definite advantage \cite{bache:2007b} because it surprisingly
gave a more distorted compressed pulse. Moreover a very large phase
mismatch was observed \cite{laegsgaard:2007}, which required a very
large effective quadratic nonlinear response by thermal poling of the
fiber.

Therefore we here investigated index-guiding microstructured polymer
optical fibers, because poling of polymers or the inclusion in the
polymer matrix of nanomaterials potentially can meet the requirement
of a large effective quadratic nonlinear response $\deff$. A definite
answer to the required nonlinearity cannot be given since at present
the investigated polymer type (Topas) has yet to have its material
Kerr nonlinear refractive index $n_{\rm Kerr}^I$ measured. However,
based on $n_{\rm Kerr}^I$-measurements of other similar polymer types
like PMMA, we estimate that it is around 10 times that of silica.
Consequently $\deff$ must exceed 5 pm/V and ideally be around 10 pm/V
if efficient generation of large negative nonlinear
phase shifts. This does pose a serious challenge for fabricating such
an mPOF, but we believe it should be possible.

GVM is considered a major obstacle for
obtaining high-quality \cite{bache:2007a} and short duration
compressed pulses \cite{bache:2008}. In order to substantially reduce
GVM we found from dispersion calculations of the transverse fiber
modes that the relative hole size had to be quite large and at the
same time the hole pitch had to be close to the wavelength of the pump
light; this is consistent with the results obtained in silica
microstructured fibers \cite{bache:2005a}.  The problem with choosing
such a design is that the SH is no longer single moded, and partial
phase matching can take place to any of the SH transverse modes,
thereby giving rise to many cascaded effects.  This is clearly
undesirable.\footnote{Note that the study in \cite{bache:2005a} in
  case of a multi-moded SH assumed that phase matching to the lowest
  order SH mode was done using QPM, so a single-moded SH
  was not as crucial as it is here.} On the other hand, the main goal of
reducing GVM was to enter the stationary regime, where efficient and
clean compression may take place. Since the fiber designs turned out
to be in the stationary regime almost irrespective the size of the
hole pitch or relative hole size, then the motivation for reducing the
GVM fades. 

The main reason why the fiber designs are always in the stationary is
the large phase mismatch. Thus, while the phase mismatch gives
problems in demanding a large $\deff$, it has the benefit of leaving
the cascaded interaction in the stationary regime, which is exactly
characterized by the phase mismatch dominating over GVM effects
\cite{bache:2007a,bache:2008}. It is worth to remark that we here
present a fiber design where compression can occur in the stationary
regime at $\lambda_1=800$ nm; most bulk quadratic nonlinear materials
are in the nonstationary regime at this wavelength \cite{bache:2007a}.

The surprising conclusion is therefore that a fiber design, where the
dispersion is not too different from the material dispersion, gave the
best compression. Numerical simulations evidenced that few-cycle
pulses at a wavelength of 800 nm can be generated from
oscillator-level pulse energies (a few nJ). Such pulse compression was
demonstrated with an endlessly single-mode mPOF with a relative hole
size of $D=0.4$ both with a large pitch $\Lambda=7.0\mic$ and a small
pitch $\Lambda=1.0\mic$. In both cases an effective nonlinearity of
$\deff=10$ pm/V was used. The only advantage of the small-pitch design
was that much lower pump powers and pulse energies are required, but on
the other hand it is much more difficult to actually draw such a
small-pitch mPOF.

A natural question is whether the polymer material can sustain the
large intensities involved when dealing with fs pulses; typically kWs
of power are focused to mode areas of a few$\mic^2$ giving intensities
ranging from $10-1000~{\rm GW/cm^2}$. However, because the fs pulses
are so short in time, the energy fluences in the examples shown in
this paper were $<10~{\rm mJ/cm^2}$, a value that unlike the
intensity does not change much during compression. Such fluences
should be below the threshold for writing, e.g., gratings in polymer
materials \cite{Baum:08}, and we therefore expect that
material damage can be avoided.

We can therefore state that if an mPOF can be drawn with large enough
effective nonlinearities to outbalance the material Kerr nonlinearity,
then only a few cm's of fiber is needed in order to generate clean
few-cycle pulses in the near-IR with very low pulse energies. This
would be a remarkable add-on to any fs pulsed oscillator. 

\section{Acknowledgements}
\label{sec:Acknowledgements}

The work presented could not have been completed without fruitful
discussion with J. Moses, F. W. Wise, J. L\ae gsgaard, and O. Bang.
Financial support from The Danish Natural Science Research Council
(FNU, grant no.  21-04-0506) is acknowledged.

\appendix

\section{Generalized propagation equations}
\label{sec:Gener-prop-equat}

For SHG in a wave-guiding medium with both quadratic and cubic
nonlinear material response, a general propagation equation for the
electrical fields including self-steepening and cubic higher order
nonlinear terms can readily be derived from
Refs.~\cite{moses:2006b,bache:2007}. There bulk propagation is
considered, but with some slight modifications that will become
apparent in what follows the propagation equations for the electrical
fields $\Es_j(z,\tau)$ read
\begin{subequations}
\label{eq:shg-bulk}
\begin{align}
  \label{eq:shg-bulk-fh-raman}
  &\Lm_1\Es_1+\kappa_{\rm SHG,1}\hat S_1\Es_1^*\Es_2e^{i\Delta \kz z}
\\&
+\kappa_{\rm Kerr,1}\hat S_1\left[\hat
  \Es_1\left(\mathcal{F}_{11}|\Es_1|^2
  +2\mathcal{F}_{12}|\Es_2|^2\right)\right]=0
\nonumber
\\  
  \label{eq:shg-bulk-sh-raman}
  &\Lm_2\Es_2+\kappa_{\rm SHG,2}\hat S_2\Es_1^2e^{-i\Delta \kz
  z}
\\\nonumber&
+\kappa_{\rm Kerr,2}\hat S_2\left[
  \Es_2\left(\mathcal{F}_{22}|\Es_2|^2+2\mathcal{F}_{12}|\Es_1|^2\right)
  \right]=0 
\end{align}
\end{subequations}
The fields are taken scalar $\E(\x,z,t)=\frac{1}{2}\hat x\sum_{j=1}^2
E_j(\x,z,t)e^{-i\omega_j t} +{\rm c.c.}  $, by assuming they are
polarized along the same polarization direction $\hat x$. The
transverse field is split from the longitudinal propagation field by
looking for solutions on the form
\begin{align}
  \label{eq:A-F-long-transverse}
  E_j(\x,z,t)=\Es_j(z,t)F_j(\x)e^{i\beta_j z}
\end{align}
where $\beta_j$ are the mode propagation wave numbers and $F_j$ the
transverse mode profiles, and $\x=(x,y)$.

The linear propagation operators are
\begin{subequations}
  \label{eq:Lprop}
\begin{align}
  \label{eq:Lprop_fh}
  \Lm_1&\equiv i\frac{\partial}{\partial z}+i\frac{\alpha}{2}
+\Dm_1, 
\\
  \label{eq:Lprop_sh}
  \Lm_2&\equiv i\frac{\partial}{\partial
    z}+i\frac{\alpha}{2}-id_{12}\frac{\partial} {\partial \tau}
  +\Dm_{2,\rm eff}
\end{align}
\end{subequations}
where $\hat D_j$ are dispersion operators up to
order $m_d$
\begin{subequations}
\begin{align}
\label{eq:Dprop}
\Dm_j&\equiv\sum_{m=2}^{m_d}i^m
  \frac{\kz_j^{(m)}}{m!}\frac{\partial^m}{\partial 
  \tau^m} 
\\
\Dm_{2,\rm eff}&\equiv \Dm_2+
\hat
  S_2^{-1}\frac{d_{12}^2}{2\beta_2}\frac{\partial^2 }{\partial\tau^2} 
\label{eq:Dsh-eff}
\end{align}
\end{subequations}
The unusual form of the SH dispersion~(\ref{eq:Dsh-eff}) is discussed
below. The mode effective indices $\neff{j}$ are related to
the propagation constants as $\kz_j=\neff{j}\omega_j/c$. The
fields are in the frame of reference traveling with the FW group
velocity $v_{g,1}$ by the transformation $\tau=t-z/v_{g,1}$, which
gives the group-velocity mismatch term $d_{12}=1/v_{g,1}-1/v_{g,2}$,
where $v_{g,j}=\bp_j$, and $\kz_j^{(m)}\equiv \partial^m
\kz_j/\partial \omega^m|_{\omega=\omega_j}$. Linear losses are
included through the loss-parameter $\alpha$. Finally, $\Delta
\kz\equiv \kz_2-2\kz_1$ is the phase mismatch.

The quadratic nonlinear coefficients are
\begin{align}
  \label{eq:kappa-bulk}
  \kappa_{{\rm SHG},j}&\equiv \frac{\omega_1}{2c\neff{j}}\frac{|\int
  d\x  \chit(\x)[F_1^*(\x)]^2F_2(\x)|}{a_j}
\\\nonumber  &
=\frac{\omega_1\deff}{c\neff{j}}\frac{|\int_P  
  d\x [F_1^*(\x)]^2F_2(\x)|}{a_j}
\end{align}
where the mode overlap area is 
\begin{align}\label{eq:modearea}
a_j\equiv \int d\x |F_j(\x)|^2,   
\end{align}
and $\chit$ is the quadratic nonlinear tensor value along the polarization
direction of the interacting waves, and $\deff\equiv \chit/2$ in the
reduced Kleinman notation. The dependence of $\chit$ on $\x$ implies
that in an index-guiding mPOF the nonlinearity is only present in the
polymer, as reminded by the subscript ``P'' in the integral of
Eq.~(\ref{eq:kappa-bulk}).

The cubic nonlinear coefficients are
\begin{align}
  \label{eq:sigma-kerr-bulk}
  \kappa_{{\rm Kerr},j}&=\frac{3\omega_j{\rm Re}(\chi^{(3)})}{8 c
  \neff{j}a_j}
=\frac{\omega_j}{ca_j}n_{{\rm Kerr},j}, 
\end{align}
where $n_{{\rm Kerr},j}\equiv 3{\rm Re}(\chi^{(3)})/8 \neff{j}$, and
$\chi^{(3)}$ is the cubic
nonlinear tensor in the polarization direction of the interacting
waves. We neglect two-photon absorption so ${\rm Im}(\chi^{(3)})=0$.
The mode overlap integrals are defined as
\begin{align}
\mathcal{F}_{jk}\equiv\int_P d\x |F_j(\x)|^2|F_k(\x)|^2 .  
\end{align}

In order to describe ultra-short pulses adequately, the usual approach
adopted in silica fibers is to divide the material Kerr response must
be divided in an electronic response and a vibrational (Raman)
response \cite{blow:1989}.  However, for the polymer materials used
for optical fibers, the delayed nature of the Raman response is
unknown. Therefore we decided to neglect the delayed vibrational Raman
response. This can be also be justified by the fact that the
propagation distances are very small, on the order of a few cm.

Finally, we have included steepening terms through a self-steepening
operator $  \hat S_j\equiv 1+\frac{i}{\omega_j}\frac{\partial}
{\partial \tau}$. 

The equations~(\ref{eq:shg-bulk}) are valid in the slowly evolving
wave approximation (SEWA)\cite{brabec:1997}, which is a general
spatio-temporal model with space-time focusing terms important for
describing fs spatio-temporal optical solitons. It was recently
extended to SHG by Moses and Wise \cite{moses:2006b}, and as a
plane-wave model for SHG with competing cubic nonlinearities by Bache
\textit{et al.}  \cite{bache:2007}. The advantage of the SEWA model is
that it does not pose any constriction on the pulse bandwidth, and
therefore holds to the single-cycle regime. Instead, the more commonly
used slowly varying envelope approximation (SVEA) only holds for
$\Delta \omega/\omega<1/3$ (and that only when including steepening
terms and the general Raman convolution response \cite{blow:1989}). In
absence of diffraction, the difference between the SHG SEWA model and
the usual SVEA model is that the SH has an \textit{effective}
dispersion term~(\ref{eq:Dsh-eff}). In the SHG SEWA model we must
remember that one assumption made when deriving
Eqs.~(\ref{eq:shg-bulk}) was that the spectra of the fundamental and
SH do not overlap (substantially). This assumption allows us to
separate the fields in two waves. We chose $\Delta
\omega/\omega_j=0.9$. This could give some overlap between the
fundamental and SH spectra, but we always made sure that the spectral
components in the overlapping regions were negligible.

We now rescale space and time (in our notation, a primed variable is
always dimensionless) so $z'\equiv z/\ld$, $\tau'=\tau/\tin$, where
$\ld\equiv \tin^2/|\bpp_1|$ is the characteristic GVD length of the FW
and $\tin$ the input pulse duration.  The fields are now normalized to
the peak input electric field $\Ein=\Es(z=0,t=0)$. Since the SH has no
input field, we choose to rescale it to the FW input field $\Ein$, so
$U_1=\Es_1/\Ein$ and $U_2=\Es_2/\sqrt{\nbar}\Ein$ , and the equations
become
\begin{subequations}\label{eq:shg-bulk-U}
\begin{align}
  \label{eq:shg-bulk-fh-U}
  \Lm_1'U_1
  &+\sqrt{|\Delta \kz'|} \nshg\hat S_1'U_1^*U_2e^{i\Delta
  \kz'z'}
\\
\nonumber
&+\nkerr^2\hat
  S_1'U_1\left[|U_1|^2+\frac{2\nbar f_{12}}
  {f_{11}}|U_2|^2\right] =0   
\\
  \label{eq:shg-bulk-sh-U}
  \Lm_2'U_2
  &+\sqrt{|\Delta \kz'|} \nshg\hat S_2U_1^2e^{-i\Delta \kz'
  z'}
\\
\nonumber
&+\frac{2\nbar^2f_{22}}{f_{11}}\nkerr^2
  \hat S_2'U_2\left[|U_2|^2+\frac{2f_{12}}
  {\nbar f_{22}} |U_1|^2\right]  =0
\end{align}
\end{subequations}
where the dimensionless SHG soliton number is defined in
Eq.~(\ref{eq:Nshg-bulk}) 
(an extension of the bulk soliton
number from Refs.  \cite{moses:2006,bache:2007} to the wave-guiding
case). The dimensionless phase mismatch is $\Delta \kz'\equiv \Delta \kz\ld$.
The cubic soliton number $\nkerr$ is given by
Eq.~(\ref{eq:Nkerr-bulk}) and is well known from the
nonlinear Schr{\"o}dinger equation (NLSE) in fiber
optics\cite{agrawal:1989}. 
In Eq.~(\ref{eq:shg-bulk-U}) the usual overlap integrals appear
\begin{align}
  \label{eq:Kerr-ovl-norm}
    f_{jk}\equiv\frac{\mathcal{F}_{jk}}{a_ja_k} =\frac{\int_P d\x
    |F_j(\x)|^2|F_k(\x)|^2 }{\int d\x |F_j(\x)|^2\int d\x|F_k(\x)|^2}
\end{align}
Finally and $\nbar \equiv
\neff{1}/\neff{2}$ which is typically close to unity. The
dimensionless propagation operators are
\begin{subequations}
\begin{align}
  \label{eq:Lprop_fh-dimless}
  \Lm_1'&\equiv i\frac{\partial}{\partial z'}+i\frac{\alpha'}{2}
+\Dm'_1,
 \\ 
  \label{eq:Lprop_sh-dimless}
  \Lm_2'&\equiv i\frac{\partial}{\partial
    z'}+i\frac{\alpha'}{2}-id_{12}'\frac{\partial} {\partial \tau'} +\Dm'_{2,\rm eff}
\\
\Dm_j'&\equiv \sum_{m=2}^{m_d}i^m \delta_j^{(m)}\frac{\partial^m}{\partial
  \tau'^m}
\end{align}
\end{subequations}
where we have introduced the dimensionless loss $\alpha'=\alpha \ld$
and the dimensionless dispersion coefficients
\begin{align}
  d_{12}'&\equiv d_{12}\frac{\ld}{\tin}
, \quad 
\delta_j^{(m)}\equiv \ld
\frac{1}{\tin^{m}m!}\kz_j^{(m)}
\end{align}
Finally, the steepening operators working with dimensionless
time are $\hat S_1'\equiv \left(1+is'\frac{\partial} {\partial
    \tau'}\right)$ and $\hat S_2'\equiv
\left(1+i\frac{s'}{2}\frac{\partial} {\partial \tau'}\right)$,
where $ s'\equiv(\omega_1\tin)^{-1}$. The SH effective
dispersion~(\ref{eq:Dsh-eff}) 
in  dimensionless form is 
\begin{align}
  \label{eq:D2-eff-dimless}
  \Dm'_{2,{\rm eff}}&\equiv\Dm_2'+\hat
  S_2'^{-1}\frac{\nu}{2}\frac{\partial^2 }{\partial\tau'^2} 
\end{align}
where the dimensionless factor
$\nu\equiv\frac{c\ld}{\omega_2 \neff 2 \lgvm^2}$. By using 
$  \hat S_2'^{-1}=\sum_{m=0}^\infty
  \left(\frac{-is'}{2}\right)^m\frac{\partial ^m}{\partial \tau'^m}$
we get\cite{bache:2007}
\begin{align}
  \label{eq:D2-eff-dimless-final-app}
  \Dm'_{2,{\rm eff}}=\sum_{m=2}^{m_d} i^m\left[
  \delta_2^{(m)}+\frac{\nu}{2}\left(\frac{s'}{2}\right)^{m-2}\right]
\frac{\partial ^m}{\partial \tau'^m}
\end{align}

The dimensionless propagation equations~(\ref{eq:shg-bulk-U}) are the
starting point of the analysis. The difference between the bulk
equations we presented in Ref.~\cite{bache:2007} is that the soliton
numbers and the coefficients for the SPM and XPM terms are modified to
include the mode overlap areas, and that we are dealing with power and
mode propagation constants instead of intensity and wave numbers.
However, the dimensionless form is general, so the scaling laws and
the critical transition points to compression found in
Ref.~\cite{bache:2007} will still hold.

\section{Calculation of the transverse fiber modes and dispersion}
\label{sec:Calc-fiber-modes}

The propagation equations describe the dynamics of the field envelope
in the $z$ propagation direction, and were found by describing the
field as in Eq.~(\ref{eq:A-F-long-transverse}). The transverse modes
$F_j(\x)$ will from this analysis have to obey a Helmholtz-type of
equation, whose solution will give the transverse eigenmodes $F_j(\x)$
and corresponding eigenfrequencies $\omega_j$ allowed by the
fiber. 

We calculated the fiber modes with the MIT Photonic-Bands (MPB)
package \cite{johnson:2001}. Each unit cell contained $n_{\rm
  C}^2=48^2$ grid points, and the super cell contained $n_{\rm
  SC}^2=7^2$ unit cells. For a given $\beta_j^{\rm
  MPB}=\beta_j\Lambda/2\pi$, the fundamental mode frequency
$\omega_1^{\rm MPB}=\omega_1 \Lambda/c$ and group velocity were first
calculated, followed by iterations of the SH until $|\omega_2^{\rm
  MPB}-2\omega_1^{\rm MPB}|<10^{-4}$.
Material dispersion, parameterized by a Sellmeier equation (see
App.~\ref{sec:Topas}), was then included using a perturbative
technique \cite{laegsgaard:2003}, whose advantage is that many
different $\Lambda$ values can be calculated perturbatively from the
MPB data (where $\Lambda$ is unity). From these modified data we may
then calculate the dispersion properties of the fiber including the
effect of material dispersion. We should mention that after the
perturbative technique is applied we get a modified set of
eigenfrequencies $\tilde\omega_j^{\rm MPB}$, and therefore the
requirement $|\tilde\omega_2^{\rm MPB}-2\tilde\omega_1^{\rm
  MPB}|<10^{-4}$ no longer holds. However, the $(\beta_j^{\rm
  MPB},\tilde\omega_j^{\rm MPB})$ data sets were afterwards converted
to dimensional form, and then fitted to a regular grid. This ensures
that the calculation of the SH dispersion was actually done at the
proper frequency $\omega_2=2\omega_1$. The higher-order dispersion
used in the numerics was calculated with a robust polynomial fitting
routine, that gave proper convergent results compared to the original
$\beta_j$-values when fitting up to 10
polynomial orders.

\begin{table}[b]
  \centering
  \begin{tabular}{c||c|c}
T [$^\circ$C] & A [$\mu{\rm m}^2$] & B\\
\hline
15 & 1.08199$\cdot 10^{-2}$ & 1.31211\\
25 & 1.09177$\cdot 10^{-2}$ & 1.30835\\
50 & 1.08080$\cdot 10^{-2}$ & 1.30139\\
75 & 1.09199$\cdot 10^{-2}$ & 1.29348\\
  \end{tabular}
  \caption{Sellmeier equation~(\ref{eq:sellmeier}) fitting parameters
    for refractive index data points in Topas grade 5013 measured in
    \cite{Khanarian:2001} at various temperatures and with wavelengths
    between 435.8 nm and 1014 nm. } 
  \label{tab:sellmeier}
\end{table}

\section{Topas Sellmeier equation and losses}
\label{sec:Topas}

We here make an accurate fit with a Sellmeier equation to the
refractive index data from \cite{Khanarian:2001} made in the visible
and near-IR. Using {\sc Mathematica} we fit to the following
single-resonance Sellmeier equation
\begin{align}\label{eq:sellmeier}
  n^2(\lambda)=1+B/(1-A/\lambda^2)
\end{align}
where $\lambda$ is the wavelength measured in $\mu$m. The measurements
were made from 15-75 $^\circ$C and with wavelengths between 435.8 nm
and 1014 nm \cite{Khanarian:2001}. The fitting parameters presented
here are more accurate than the ones
reported in \cite{Khanarian:2001} (could be because in
\cite{Khanarian:2001} a 3-parameter fit was used, while we use a
2-parameter fit).

\begin{figure}[tb]
  \begin{center}
\includegraphics[width=8cm]{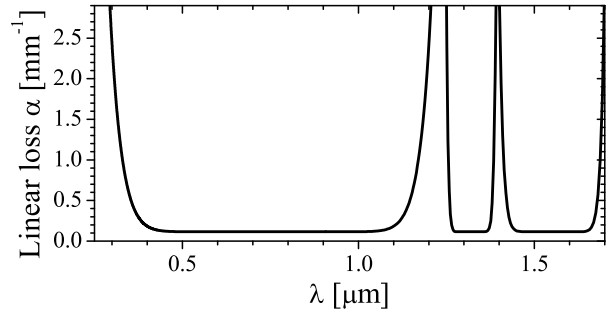}    
  \end{center}
\caption{\label{fig:loss} The loss-parameter $\alpha$ used in the
  numerics: both linear losses (estimated to 0.5 dB/cm
  \cite{Khanarian:2001}) and the absorption peaks of
  Topas, see \cite{Khanarian:2001}, are modeled. }
\end{figure}

The linear losses were included in the numerical simulations, but had
only little influence since the fiber lengths considered here were on
the order of a few cm's. The absorption peaks of Topas around
$\lambda=1.2\mic$ and $\lambda=1.4\mic$ were modeled using a loss
method: top-hat transmission profiles with a maximum transmission of
unity were fitted to the three main spectral windows
$\lambda\in[0.29,1.21]\mic$, $\lambda\in[1.25,1.35]\mic$ and
$\lambda\in[1.4,1.7]\mic$ as measured in \cite{Khanarian:2001} in a
$L=3.2$ mm sample. The linear losses were then calculated as
$\alpha=-{\rm ln}(T)/L+\alpha_0$, where $T$ is the top-hat transmission
profile, and $\alpha_0$ are the base linear losses (estimated to 0.5
dB/cm \cite{Khanarian:2001}, i.e.  $\alpha_0=0.115~{\rm
  cm^{-1}}$). The loss profile of the simulations is shown in
Fig.~\ref{fig:loss}.


\end{document}